\documentclass[aps,twocolumn,prl,showpacs,superscriptaddress]{revtex4}
\usepackage{graphicx}
\usepackage{amsmath,amssymb,latexsym,color,amsfonts}
\bibstyle{apsrev.bib}

\newcommand{\beq}{\begin{equation}}
\newcommand{\eeq}{\end{equation}}
\begin{document}


\title{Impact ionization in InSb probed by THz-pump THz-probe spectroscopy}

\author{Matthias C. Hoffmann}
\affiliation{Massachusetts Institute of Technology}
\email[mch@mit.edu]{}

\author{J\'{a}nos Hebling}
\affiliation{Department of Physics,University of P\'{e}cs,
Hungary}
\author{Harold Y. Hwang}
\author{Ka-Lo Yeh}
\author{Keith A. Nelson}
\affiliation{Massachusetts Institute of Technology}

\date{\today}

\begin{abstract}
Picosecond carrier dynamics in indium antimonide (InSb) following
excitation by below-bandgap broadband far infrared radiation were
investigated at 200 K and 80 K. Using a novel THz-pump/THz-probe
scheme with pump THz fields of 100 kV/cm and an intensity of 100
MW/cm$^2$, we observed carrier heating and impact ionization. The
number of carriers produced exceeds $10^{16}$cm$^{-3}$,
corresponding to a change in carrier density $\Delta N/N$ of 700\%
at 80K.  The onset of a well defined absorption peak at 1.2 THz is
an indication of changes in  LO and LA phonon populations due to
cooling of the hot electrons.

\end{abstract}


\maketitle

Hot electron dynamics and carrier multiplication in semiconductors
play important roles in the design of fast electronic devices as
well as in the development of highly efficient solar cells
\cite{werner:1028}. The elucidation of carrier dynamics on the
ultrashort timescale is hence of great technological as well as
fundamental interest. Indium antimonide (InSb) is a model system
for the study of hot electron dynamics due to its low bandgap of
170 meV at room temperature \cite{littler1985} and the fact that
it has the highest electron mobility and saturation velocity of
all known semiconductors. Its high mobility allows the fabrication
of transistors with extremely high switching speed,
 \cite{ashley2004} but further technological applications are
complicated by the low impact ionization threshold.  Impact
ionization by high electric fields is a well known phenomenon in
InSb \cite{ancker1972} and has been studied mostly in the quasi DC
limit of $\omega \tau\ll1$ where $\tau$ is the characteristic
electron momentum relaxation time, typically of the order of a few
picoseconds. In this limit, the extent of impact ionization is
determined by the probability of an electron gaining enough energy
from the driving field to cross the ionization threshold. An
equilibrium between energy gain of the electrons due to the
accelerating field and energy loss due to phonon scattering is
reached, leading to saturation in the drift velocity $v_d$.
Carriers with energies greater than the bandgap $\varepsilon_g$
can produce new electron-hole pairs through the impact ionization
process, which also can be viewed as an inverse Auger effect
\cite{keldysh1965}. The newly generated carriers can be
accelerated by the field and can induce impact ionization
themselves. Carrier relaxation dynamics in the high-frequency
limit ($\omega\tau\gg1$) have been studied in InSb by sub
picosecond time-resolved mid-infrared nonlinear spectroscopy using
above-bandgap excitation  \cite{lobad2004}. Below-bandgap
excitation by CO$_2$ lasers also has been used for time resolved
measurements with nanosecond resolution \cite{nee1978} and for
intensity-dependent transmission measurements
\cite{ganichev1986,ganichev1994}. Below-bandgap excitation allows
study of bulk hot-carrier dynamics without optical generation of
holes or strong absorption and high carrier densities very close
to the surface.

In this letter, we report experimental observations of carrier
generation  in InSb at 80K and 200K due to impact ionization
induced by below-bandgap IR radiation on the picosecond time
scale, where $\omega\tau\approx1$. Near single-cycle pulses with
fields strengths up to 100 kV/cm and a duration of 1 ps were used.
The rise time of the THz pulses was less than the electron
momentum relaxation time of $\tau=2.5$ ps in InSb at 77 K. This
leads to highly accelerated carriers that can cause carrier
multiplication through impact ionization
\cite{hoffmann2008,wen2008}. THz radiation also can be used as a
very sensitive \emph{probe} to directly monitor free carrier
behavior in semiconductors \cite{schmuttenmaer2000}. The
combination of sub-bandgap direct excitation of doped
semiconductors and time resolved spectroscopy provides an
excellent tool for observing carrier dynamics
\cite{hoffmann2008,hebling2008ultrafast,hebling2008}.

\begin{figure}
 \scalebox{0.45}{\includegraphics{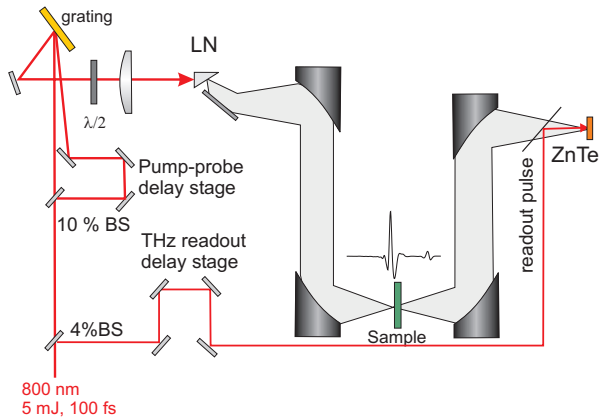}}
 \caption{The experimental setup uses pump and probe THz pulses in a collinear geometry,
 generated by tilted pulse front excitation in LiNbO$_3$  (LN)\label{setup}}
 \end{figure}

The experimental setup shown in Figure \ref{setup} was used for
collinear THz-pump/THz-probe measurements. We generated
single-cycle THz pulses by optical rectification of a 800 nm, 5.5
mJ pulse from a Ti:Sapphire laser with 100 fs pulse duration at a
repetition rate of 1 kHz. We tilted the pulse intensity front with
a grating-lens combination to achieve noncollinear velocity
matching in lithium niobate \cite{hebling2002,feurer2007},
yielding THz pulses with energies up to 3 $\mu$J
\cite{yeh2008,yeh2007}.
 The optical beam was split using a 10:90 beamsplitter into two parts
that were recombined under a small angle at the same spot on the
grating. The 10\% part was modulated using an optical chopper and
was used to generate the THz probe. The 90 \% part was variably
delayed and was used to generate the THz pump pulse. The
single-cycle THz pulses were focused onto the sample using a
90-degree off-axis parabolic mirror pair with 190 and 75 mm focal
lengths. The focused size was determined to be 1 mm using a razor
blade scan. Another off-axis parabolic mirror pair with focal
lengths of 100 and 190 mm was used to image the sample plane onto
the ZnTe detector crystal for electro-optic sampling of the THz
field using balanced detection and a lock-in amplifier
 \cite{zhang1995}. In order to ensure the linearity of the detected
signal and to eliminate THz pulse reflections
\cite{turchinovich2007}, the ZnTe sampling crystal had an active
layer of 0.1 mm and a total thickness of 1.1 mm. Selective
chopping of the probe beam provided excellent suppression of the
pump pulse. Spectral analysis of our THz pump-probe results was
conducted in the 0.2 to 1.6 THz range where the spectral amplitude
was sufficiently high. A pair of wiregrid polarizers was used to
attenuate the THz pulses for intensity-dependent studies. The
samples were a n-type Te-doped and a nominally undoped InSb wafer,
each 450 $\mu$m thick, with carrier concentrations at 77 K of
$2.0\times10^{15}$cm$^{-3}$ and $2.0-4.9\times10^{14}$cm$^{-3}$
respectively. The mobility as specified by the manufacturer was
$2.5\times10^{5}$cm$^{2}$/Vs. The THz fields were polarized
parallel to the (100) axes of the crystals. We measured the THz
fields $E(t)$ that reached the ZnTe crystal with and without the
sample in the beam path from which we calculated the effective
absorption coefficient
\begin{equation}
\alpha_{\mathrm{eff}}=-\frac{1}{d}\ln\left(T^2\cdot\frac{\int_0^{t_{\mathrm{max}}}
E^2_{\mathrm{sam}}(t)dt}{\int_0^{t_{\mathrm{max}}}E^2_{\mathrm{ref}}(t)dt}\right)
\end{equation}
where $d$ is the sample thickness, $t_{\mathrm{max}}$ is the time
window of the measurement and $T$ is a factor accounting for
reflection losses at the sample surfaces. The quantity
$\alpha_{\mathrm{eff}}$ is equivalent to the energy absorption
coefficient averaged over our bandwidth.
 \begin{figure}
 \scalebox{0.9}{\includegraphics{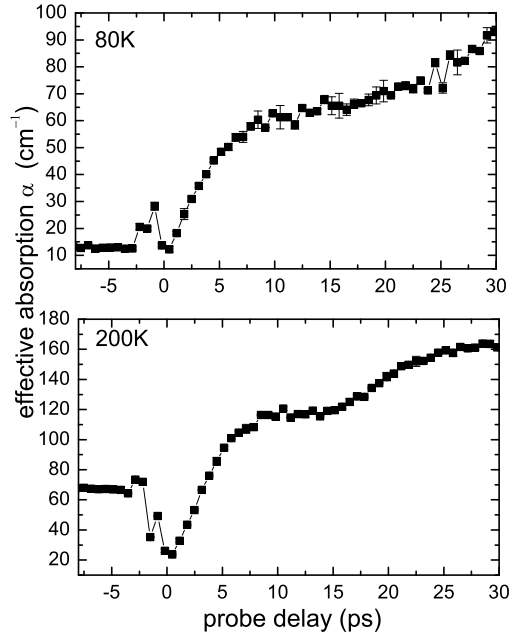}}
 \caption{THz-pump/THz-probe time-resolved absorption data, spectrally averaged over the 0.2-1.6 THz range, from doped  InSb ($N_c=2.0\times10^{15}$cm$^{-3}$) at 80 K and 200 K.
 The deviation starting at 10 ps is due to re-heating caused by the THz pulse reflection within the sample. \label{pp_both}}
 \end{figure}
Figure \ref{pp_both} shows the time-resolved absorption traces for
the doped sample at 80 K and 200 K covering a probe delay range of
up to 30 ps. At both temperatures, the absorption increases after
THz excitation and reaches a plateau after 30 ps, with a total
increase in absorption of 80-90 cm$^{-1}$. This rise is caused by
the generation of new carriers through impact ionization. Unique
to the measurement at 200 K is the initial dip in the absorption
immediately after the THz excitation.  The drop in absorption in
this case is caused by a decrease in the mobility of the highly
energetic hot electrons as a result of both the strong
non-parabolicity of the $\Gamma$ valley in the conduction band of
InSb \cite{weng1995} and scattering of these hot electrons into
side valleys
 \cite{hebling2008ultrafast,hebling2008,keilmann1986,constant1985}.
The carrier mobility $\mu$ directly influences the absorption
coefficient $\alpha$ which is typically conceptualized by the
Drude model, parametrized by the  plasma frequency
$\omega_p=(Ne^2/\epsilon_0\epsilon_\infty m^*)^{1/2}$ and the
momentum scattering rate $\gamma$ in the form
\begin{equation}
\alpha=\frac{\epsilon_\infty \omega_p^2\gamma}{n c
(\omega^2+\gamma^2)},
\end{equation}
where $\epsilon_\infty$ is the high-frequency limit of the
dielectric function and $n$ is the refractive index.
 In the low-frequency limit ($\omega\ll\gamma$) the
absorption $\alpha$ is directly proportional to the carrier
mobility $\mu=e/\gamma m^{*}$. Due to the small bandgap of InSb,
the intrinsic carrier concentration at 200 K is much higher than
that at 80 K. The effect of absorption saturation is thus much
stronger at 200 K. After 4 ps, the additional absorption caused by
the newly generated carriers exceeds the saturation effect, the
magnitude of which is also diminishing as a result of the cooling
of the hot carriers, thereby leading to the delayed rise observed
in the overall absorption.

Comparison between the equilibrium absorption of InSb at 80 K and
the value measured 30 ps after intense THz excitation shows a
eight-fold increase, indicating a similar increase in carrier
concentration from 2$\times10^{15}$ cm$^{-3}$ to
$1.5\times10^{16}$cm$^{-3}$. The same analysis cannot be applied
reliably at 200 K due to the very large intrinsic carrier
absorption, which overwhelms the dynamic range of our spectrometer
system. This effect is especially acute at low frequencies where
the absorption is strongest, leading to an apparent saturation of
the frequency- averaged absorption shown in Figure \ref{pp_both}b.
Some signal at $t<0$ appears in Fig.  \ref{pp_both}a because of
the nonlinear interaction in the LN crystal between the THz pump
and probe fields and the optical pulses that generate them as well
as nonlinear interactions in the sample.

We employ a simple system of rate equations, displayed below, to
model the dynamics of impact ionization in the first 30 ps,
assuming quadratic scaling \cite{keldysh1965} of the impact
ionization probability above the threshold energy
$\varepsilon_{th}\approx\varepsilon_g$.
\begin{eqnarray}
\frac{{dN}}{{dt}} = C(\varepsilon (t) - \varepsilon _{th} )^2
\cdot N(t) \cdot \Theta (\varepsilon (t) - \varepsilon _{th} )\\
\frac{{d\varepsilon }}{{dt}} =  - C(\varepsilon (t) - \varepsilon
_{th} )^2  \cdot \varepsilon _{th}  \cdot \Theta (\varepsilon (t)
- \varepsilon _{th} ) - \frac{\varepsilon}{{\tau _\varepsilon  }}
\end{eqnarray}
In this model $N(t)$ is the electron concentration,
$\varepsilon(t)$ is the average carrier energy and $\Theta(t)$ is
the Heaviside step function. We used a numerical value of
$C=7\times10^{50}J^{-2}s^{-1}$ obtained from Ref.
\cite{devreese1982}. The energy relaxation time $\tau_e$ was
assumed to be time and energy independent.

\begin{figure}
 \scalebox{0.8}{\includegraphics{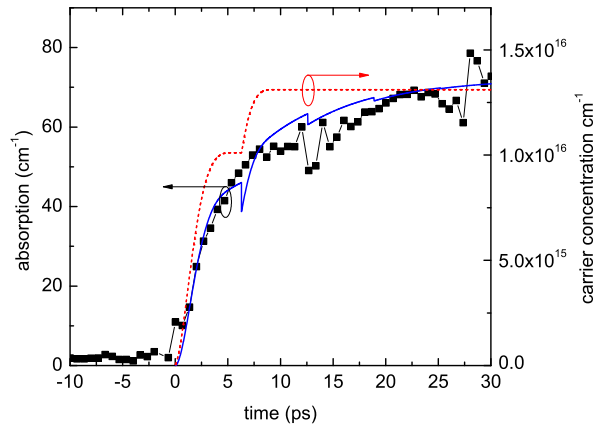}}
 \caption{Experimental data for the undoped sample (squares) and simulation results for the carrier concentration (dashed) and absorption (solid line) based on quadratic scaling of impact ionization rate with carrier energy.
 Parameters used were: $N(t=0)=5\times10^{13}$ cm$^{-3}$, energy relaxation time $\tau_e=7$ ps and
$\varepsilon_0=\varepsilon(t=0)=1.3$ eV.}\label{simulation_janos}
\end{figure}

A numerical solution, taking into account the reflection at the
sample interface, is shown in Fig. \ref{simulation_janos}. The
effect of absorption saturation due to carrier heating
\cite{hebling2008ultrafast,hebling2008} was accounted for
approximately by assuming $\alpha_{eff}\propto N(t)[\varepsilon _0
- \varepsilon (t)]/\varepsilon _0 $ where $\varepsilon_0$ is the
average carrier energy immediately after excitation. From this fit
we obtain a phenomenological relaxation time $\tau_\varepsilon$
time of 7 ps, much longer than the 1.3-2 ps calculated for the
energy relaxation time in the DC limit by Kobayashi
\cite{kobayashi1977}. The faster value cannot describe our
experimental conditions since it would preclude impact ionization
and carrier cooling dynamics that we observe at slower time
scales.
\begin{figure}
\scalebox{0.9}{\includegraphics{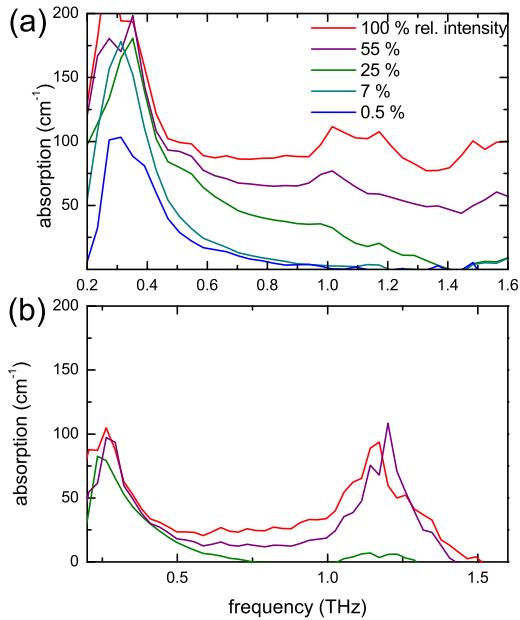}} \caption{THz absorption
spectra at various pump intensities measured at a probe delay time
of 35 ps for the doped sample (a) and undoped sample (b) at 80K}
\label{intdep}
\end{figure}
In order to elucidate the lattice dynamics further, we performed a
series of intensity-dependent pump-probe measurements with a fixed
probe delay. Figure \ref{intdep} shows the absorption spectra
obtained with different pump intensities at a probe delay of 35 ps
 for the doped and undoped samples at 80K. At
frequencies below 0.6 THz, we observe the expected Drude-like
contribution from free carrier absorption which is more pronounced
at higher pump fluence. In addition, we observe a distinct
absorption peak at 1.2 THz in the undoped sample and a weak
feature that indicates a similar peak in the doped sample. The
amplitude of the peak is highly intensity dependent, and appears
to approach its asymptotic value just above 50\% of the maximum
intensity. The behavior of this peak suggests that its origin is
lattice vibrational rather than electronic. Polar optical phonon
scattering is well known as the dominant energy loss mechanism for
hot electrons in InSb  \cite{conwell1967}. The main channel of
energy loss is through the emission of LO phonons with a frequency
of 5.94 THz. These phonons decay into acoustic modes through
anharmonic terms of the crystal potential and through the
second-order electric moment of the lattice \cite{ferry1974}. A
series of sum- and difference phonon peaks between 1 to 10 THz has
been observed and assigned in the far-infrared spectrum of InSb
 \cite{koteles1974}. At very low THz fields, produced by a
photoconductive antenna, we also were able to observe some of
these weak absorption peaks in the undoped sample. The assignments
reported in \cite{koteles1974} indicate a 1.2 THz difference
frequency between the LO and LA mode at the zone boundary.
 The drastic
change in the absorption coefficient of the difference phonon peak
is the result of large changes in phonon populations due to energy
transfer from the hot electrons generated by the THz pump pulse.
Monte-Carlo simulations \cite{brazis2008} have shown that
substantial phonon population changes can occur even at
comparatively low DC fields on picosecond timescales.
 \begin{figure}
 \scalebox{1}{\includegraphics{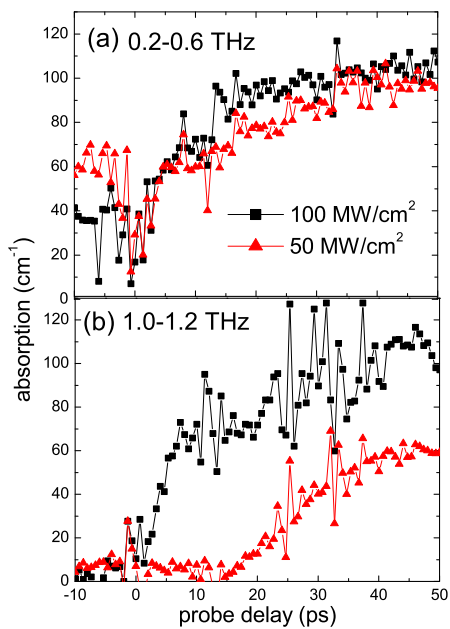}}
 \caption{Time-resolved spectrally averaged absorption for the
 doped sample at 80 K with 100\% pump intensity (black squares) and 50\% pump intensity (red triangles).
 (a) the average absorption between 0.2 and 0.6 THz, (b) the average absorption between 1.0-1.2 THz}
 \label{abspowerdep}
 \end{figure}
The temporal evolution of the electron-lattice interaction can be
studied by separating the spectrally resolved pump probe data into
two different frequency bands 0.2-0.6 THz and 1.0-1.2 THz.
  This is illustrated in Figure \ref{abspowerdep} where data from time-resolved
measurements at full THz intensity and at half the full intensity
are shown for the doped sample. For the low-frequency band, the
rise in absorption is almost identical at the two intensities,
while the contribution from the absorption between 1.0 to 1.2 THz
is reduced considerably when the THz intensity is halved. The
higher frequency portion of the absorption spectrum is clearly
more sensitive to the intensity of the THz pump pulse, as
illustrated in Figure \ref{intdep}b. In addition, at the lower
intensity there is a delay of approximately 10 ps before the rise
of the absorption signal. This delay was reproduced qualitatively
for all measurements at intermediate THz pump levels. We do not
have a complete understanding of the intensity-dependence of the
delay, but we note that the delay is on the order of the LO phonon
decay time in InSb \cite{ferry1974}, suggesting that  it arises
from electronic and LO phonon relaxation processes that populate
the LA phonons. The large fluctuations observed at 12 and 24 ps
are due to THz pump pulse reflections in the sample that are
overlapped in the EO crystal with the THz probe pulse.

 The
newly developed THz pump/THz probe technique permits sensitive
monitoring of carrier dynamics in semiconductors on the picosecond
timescale. We observed the dynamics of impact ionization and
carrier generation, up to a seven-fold increase over the
equilibrium carrier-generation, following intense THz excitation
of InSb. Our ability to spectrally analyze the time-resolved
signal allowed us to detect distinct features that we attribute
decay of electronic energy into LO and LA modes. Monte-Carlo
simulations are needed for a more in-depth understanding of the
interplay between hot electrons and lattice, taking into account
effects like impact ionization, intervalley- and polar optical
phonon scattering and changes in phonon population. Additional
effects of the THz fields themselves, including the possibility of
THz-induced band-to-band tunneling\cite{kane1959} that could
produce new carriers directly, also warrants further analysis.

\begin{acknowledgments}
We would like to thank R. Brazis for stimulating discussions. This
work was supported in part by ONR grant no. N00014-06-1-0463.
\end{acknowledgments}

\bibliography{insb}

\end{document}